# Deformation of μm- and mm-sized Fe2.4wt.%Si single- and bi-crystals with a high angle grain boundary at room temperature

M. Heller[a], J. S. K.-L. Gibson[a], R. Pei[a], S. Korte-Kerzel[a]

[a]*Institute of Physical Metallurgy and Materials Physics, RWTH Aachen University, Aachen, Germany*

## Abstract

Plasticity in body-centred cubic (BCC) metals, including dislocation interactions at grain boundaries, is much less understood than in face-centred cubic (FCC) metals. At low temperatures additional resistance to dislocation motion due to the Peierls barrier becomes important, which increases the complexity of plasticity. Iron-silicon steel is an interesting, model BCC material since the evolution of the dislocation structure in specifically-oriented grains and at particular grain boundaries have far-reaching effects not only on the deformation behaviour but also on the magnetic properties, which are important in its final application as electrical steel. In this study, two different orientations of micropillars (1, 2, 4 μm in diameter) and macropillars (2500 μm) and their corresponding bi-crystals are analysed after compression experiments with respect to the effect of size on strength and dislocation structures. Using different experimental methods, such as slip trace analysis, plane tilt analysis and cross-sectional EBSD, we show that direct slip transmission occurs, and different slip systems are active in the bi-crystals compared to their single-crystal counterparts. However, in spite of direct transmission and a very high transmission factor, dislocation pile-up at the grain boundary is also observed at early stages of deformation. Moreover, an effect of size scaling with the pillar size in single-crystals and the grain size in bi-crystals is found, which is consistent with investigations elsewhere in FCC metals.

## 1. Introduction

Iron-silicon sheet steel is the most widely used material for the iron cores of electrical machines like motors, generators and transformers. With the trend of thinner and thinner sheets to minimise the eddy current losses [1], the grain size is approaching the sheet thickness. This makes the response of individual orientations of single grains and grain boundaries important for the final shear cutting as well as for the evolving magnetic properties. The latter are negatively affected by lattice obstacles such as dislocation tangles and pile ups [2]. Thus, a better understanding of the evolving dislocation structures at grain boundaries can help to ultimately improve these properties. In this study, we aim to contribute to this topic by studying the effect of an individual grain boundary in Fe2.4wt.%Si on the dislocation structures formed during deformation. For this, we use compression tests of single- and bi-crystalline samples at the micrometre and millimetre scale analysed by (scanning) electron microscopy and electron backscatter diffraction.

The emergence of micro- and nanomechanical test methods [3-5] enables the systematic investigation of various physical phenomenon that previously was only possible through laborious study, such as the influence of individual grain boundaries on the deformation behaviour [6-14]. Most samples for these experiments are milled with a focused ion beam (FIB), which allows the milling of bi-crystalline micropillars with carefully positioned grain boundaries [6-21].

Until now, most studies focused on single- and bi-crystalline FCC nano- to micropillars [6-14, 16-23] and to our knowledge just one study by Weaver et al. [15] exists dealing with a bi-crystalline BCC metal. Dislocation motion in BCC metals is much less straightforward than in FCC metals. A variety of slip planes are observed, both on the expected planes with a large interplanar spacing ({110}, {112}, {123}) but also on maximum resolved shear stress planes (MRSSP) as well as anomalous slip [24-27]. The greatest contrast to FCC metals is the influence of the high Peierls barrier in the deformation in BCC materials at temperatures below the critical temperature ($T_c$) [28, 29], resulting in plasticity being limited by the double-kink nucleation of screw dislocations [25, 30]. Above $T_c$, other mechanisms like work-, precipitation- or solid solution hardening outweigh the strength contribution of the Peierls barrier and an athermal regime begins. For α-iron the critical temperature is around room temperature [31]. Thus, it is not *a priori* clear which mechanism dominates during the room-temperature experiments conducted here. A further complexity of slip in BCC metals is associated with the non-planar core structure of the screw dislocations leading to non-Schmid behaviour [25]. The activation of a specific slip system is therefore dependent on temperature, purity, strain rate, the dislocation core structure, the controlling hardening mechanism and the loading conditions [25].

With respect to the dislocation mechanisms during the deformation of bi-crystals, it is of major interest whether dislocations transmit [32], pile up [33] or annihilate [34] at a grain boundary or whether the grain boundary might act as a dislocation source [35]. All these mechanisms can of course be active at the same time depending on which dislocation interactions are present. Thus, it is important to quantify which mechanism dominates and under which conditions. The transmission factor *m'* (Equation 1) as proposed by Luster and Morris [36] can give a first hint on what will likely happen at a grain boundary. *m'* can take values between 0 and 1 and is calculated according to:

$$m' = cos(\psi)\,cos(\kappa) \tag{Equation 1}$$

where ψ is the angle between the slip plane normals and κ the angle between the slip directions of the considered slip systems. For high values of *m'* there is only a small deviation between the selected Burgers vectors in adjacent grains during deformation, theoretically leading to an easy transmission through the grain boundary. Weaver et al. [15] found direct slip transmission in BCC bi-crystalline micropillars of pure tantalum for *m'* ≥ 0.85 and no transmission for *m'* ≤ 0.46. This is in good agreement with the work of Bieler et al. [37] who found direct slip transmission for *m'* ≥ 0.85 and no transmission for *m'* ≤ 0.6 in the same material but in poly-crystalline, macroscopic samples.

The influence of specific grain boundaries on both the scaling of the yield stress with specimen size, as well as the activation of slip systems during microcompression, is also not yet fully understood in BCC metals. Most recent studies show a comparable size effect in bi-crystals relative to single-crystals, however, bi-crystals appear to scale with their grain size instead of the sample diameter [7, 8]. One exception is represented by the coherent Σ3 twin boundary in copper, which has no influence on the mechanical response compared to its single-crystalline counterparts, as identical Burgers vectors are present in both grains [6]. Another exception was presented by Kheradmand et al. [21] who found lower stresses in bi-crystals indicating that the grain boundary can act as a dislocation source.

This study aims to contribute to the fundamental understanding of plasticity at grain boundaries in BCC metals over different size scales. The ultimate goal is then to provide a physical basis for the correlation of the electromagnetic performance of (locally) deformed electrical sheets with the underlying dislocations stored during the final cutting process, before the final lamination and assembly of electrical steel parts takes place [38].

## 2. Experimental Methods

### 2.1. Samples

All micropillars were milled on one macroscopic, bi-crystalline, disc shaped sample ($d$ = 16 mm) (Figure 1). This sample in turn was taken from a cylindrical, macroscopic bi-crystal grown at 0.1 mm/min according to the vertical Bridgman-Stockbarger method, using high-purity iron alloyed with Fe65wt.%Si to give Fe2.4wt.%Si. Discs for micro- and macropillars were cut from the grown crystal by electrical discharging machining (EDM). The flat surface finish was achieved by a final polishing with OP-U suspension (Struers).

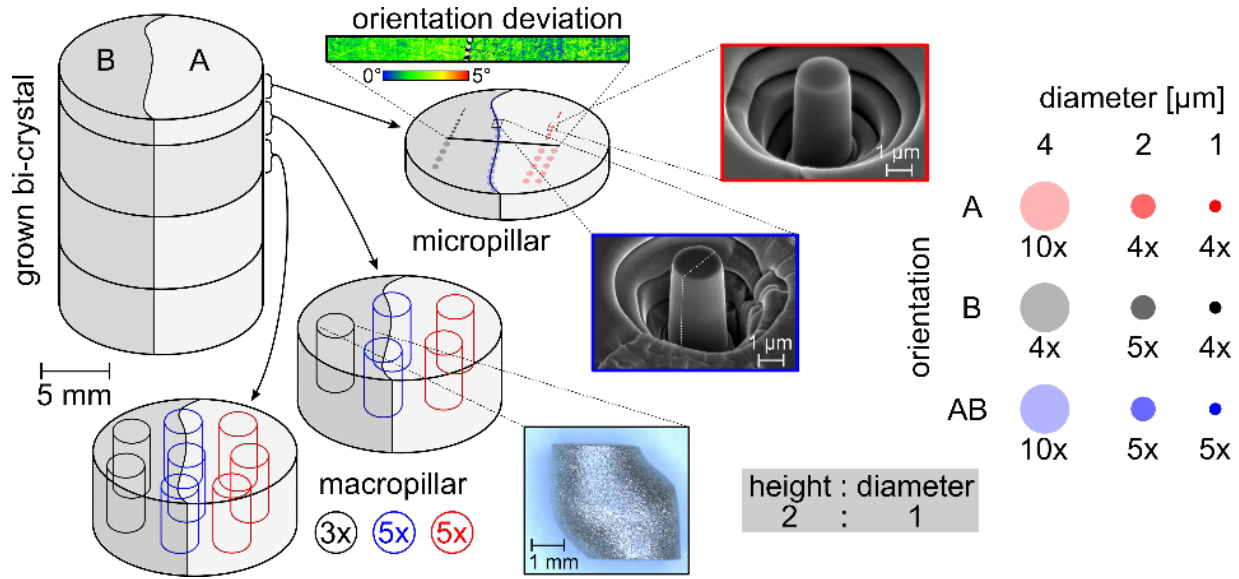

*Figure 1: Overview showing the diameter, kind (single-, bi-crystal), colour code and quantity of the different pillars as well as the orientation deviation from the mean orientation (misorientation deviation map) of the grain along the disk shaped sample.*

## 2.2. Micro- and Macropillars

Figure 1 visualises the experimental matrix of this study. All experiments were carried out on two single-crystal orientations (A and B) and their bi-crystalline counterpart (AB). In Bunge Euler Angles ($\varphi_1$, $\Phi$, $\varphi_2$), orientation A can be written as (45.4°, 7.7°, 319.7°) and B as (178.9°, 39.9°, 166.4°) resulting in a mixed ~49.2° high angle grain boundary rotated around the rounded $(\bar{6}31)$ direction.

Three micropillar sizes (1, 2 and 4 µm diameter) and one reference macropillar size (2500 µm diameter) were prepared with one to ten pillars per parameter combination (crystal orientation(s) and size). Fewer macroscopic tests were carried out because of material scarcity and proven reliability, but at least four successful tests were conducted per micropillar category. All micropillars were prepared using site-specific FIB milling (Helios NanoLab 600i, FEI Co.) utilising a final beam current of 80 pA. Due to the extensive testing scheme (51 micropillars) a simple cylindrical micropillar geometry with an aspect ratio of two was chosen to enable fast milling despite a few degrees of final taper (2-4°). The grain boundary in the AB micropillars was positioned as centrally as possible and ran perpendicular to the top of the pillar in most cases. The resulting mean geometries and standard deviations for the different pillar groups are summarized in the supplementary material (S2).

## 2.3. Compression

All microscopic compression tests were carried out on a load controlled in-situ nanoindenter (InSEM HT, Nanomechanics Inc.) utilising a 6 µm diamond flat punch. To avoid any effect of a feedback loop, a load controlled method was chosen with load rates between 0.0063 mNs$^{-1}$ and 0.4 mNs$^{-1}$ scaled by cross-sectional area, resulting in the desired strain rate of ~0.0001 s$^{-1}$ within the elastic regime. The load-displacement data were corrected for frame stiffness, thermal drift, spring deflection, and elastic deformation of the surrounding material via a Sneddon correction [39]. To calculate the engineering stresses, the top diameter was used [40]. The yield point was taken at a plastic strain of 0.5 % to filter out noise related strain jumps. The majority of the tests were carried out to a maximum strain of 25-30 % to obtain pronounced slip (hereafter referred to as "highly-deformed" pillars), while some were stopped at 5-8 % strain to check whether different slip systems are activated as deformation progresses, and are referred to as "lightly-deformed" pillars.

All macroscopic compression tests were carried out in air at room temperature at the Institute of Metal Forming (IBF, RWTH Aachen University) on a deformation dilatometer (DIL 805, TA Instruments). Deformation was strain-rate controlled with a rate of 0.01 s$^{-1}$ and a maximum strain of 30 %.

2.4. Slip system analysis

Figure 2 shows the slip system analysis approach. Micropillars were imaged from at least two positions to give an overview at 45° tilt before and after compression. All images were taken on a FEG-SEM (Helios NanoLab 600i, FEI Co.) using a through the lens SE detector at a working distance of 2-4 mm, an acceleration voltage of 10 kV and a current of 0.69 nA. Orientation data was obtained by electron backscatter diffraction (EBSD) (Hikari, EDAX Inc.) at 20 kV and 1.4 nA with a 0.3-0.5 µm step size.

Crystal Maker (CrystalMaker Software Limited) was used to visualise BCC unit cells to compare the theoretical and measured glide plane angle. Moreover, a MATLAB script utilising the toolbox MTEX [41] was written and used to visualise the slip systems in a rotatable, cylindrical model, to give a more direct comparison to the experimental micropillars (Figure 2 (c)). Geometrical transmission factors were calculated for different sets of activated slip systems using the MATLAB toolbox STABiX v3 [42].

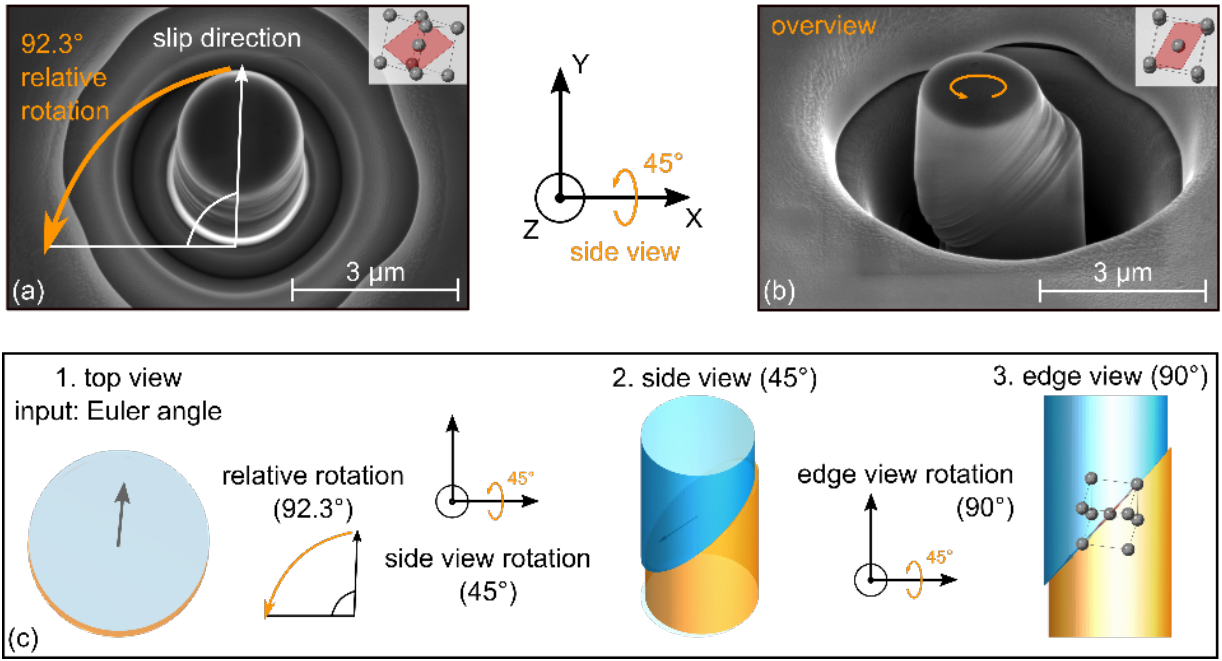

*Figure 2: Slip system analysis using direct plane angle measurements: (a) Top view with marked relative rotation to obtain the horizontal projection of the slip direction, (b) SE-image at 45° tilt in side view (at least two per pillar), (c) Slip system analysis using comparison with theoretical deformation for a given slip system with a 0°/45°/90° tilted view before and after relative rotation.*

## 2.5. Cross section EBSD

Selected micropillars were further analysed after deformation using cross section EBSD. For this, micropillars were lifted out of the disc sample onto a TEM grid, then thinned using the FIB to the centre of the pillar. EBSD measurements were then performed on the cross-section with a step size of 50 nm at 15 kV and a vacuum level of $1\times10^{-7}$ mbar or better. All collected EBSD data was post processed using OIM Analysis$^{TM}$ (EDAX Inc.).

## *3.* Results

### 3.1. Slip System Analysis

#### 3.1.1. Single-Crystal

In Figure 3, two different micropillars of orientation A with 1 µm (a) and 2 µm (b) diameter are displayed. For 1 µm - grain A the slip system $(21\bar{3})[\bar{1}\bar{1}1]$ is found to give the best fit and is therefore thought to be the active system, and is shown overlaid in black on the right-hand SE-image. This slip system was found in about one sixth of the A micropillars and has the fourth highest Schmid factor (supplementary material S3). For comparison, other high-Schmid factor slip systems are also overlaid, ordered by decreasing Schmid factor from top (black) to bottom (white). To obtain the correct horizontal width of the overlaid slip traces, the cylindrical micropillar model was adjusted according to the diameter of the micropillar and the apex of the slip trace, as seen under a 45° stage tilt, was chosen as a starting point for the overlaid lines.

The second arrow in (a) illustrates that more than one slip system can be activated in A micropillars. Here, $(1\bar{1}2)[1\bar{1}\bar{1}]$ is the matching slip system which is also the result in the analysis of the micropillar shown in Figure 3 (b). $(1\bar{1}2)[1\bar{1}\bar{1}]$ is activated in the majority of A micropillars and has the highest Schmid factor (supplementary material S3).

The two active slip systems in double-slipping pillars are always the same combination of a primary $(21\bar{3})[\bar{1}\bar{1}\bar{1}]$ slip system and a secondary $(1\bar{1}2)[1\bar{1}\bar{1}]$ system. However, the chance of double slip occurring is size dependent: ~40% of 1 µm micropillars, ~10% of 2 µm micropillars and zero 4 µm micropillars show double slip.

With increasing micropillar size, slip traces become less sharp but more of them appear. As seen in Figure 3, there are only two slip traces visible in the 1 µm micropillar in (a) and several of varying width in the 2 µm micropillar in (b). In isolated cases, the activated slip systems in orientation A were found to correspond to $(1\bar{2}3)[1\bar{1}\bar{1}]$ with a Schmid factor of 0.488 (two cases) and $(2\bar{1}3)[1\bar{1}\bar{1}]$ with a Schmid factor of 0.491 (one case). The cylindrical micropillar models in-between the SE-images each show the best matching slip system, with the arrow indicating the assumed slip direction.

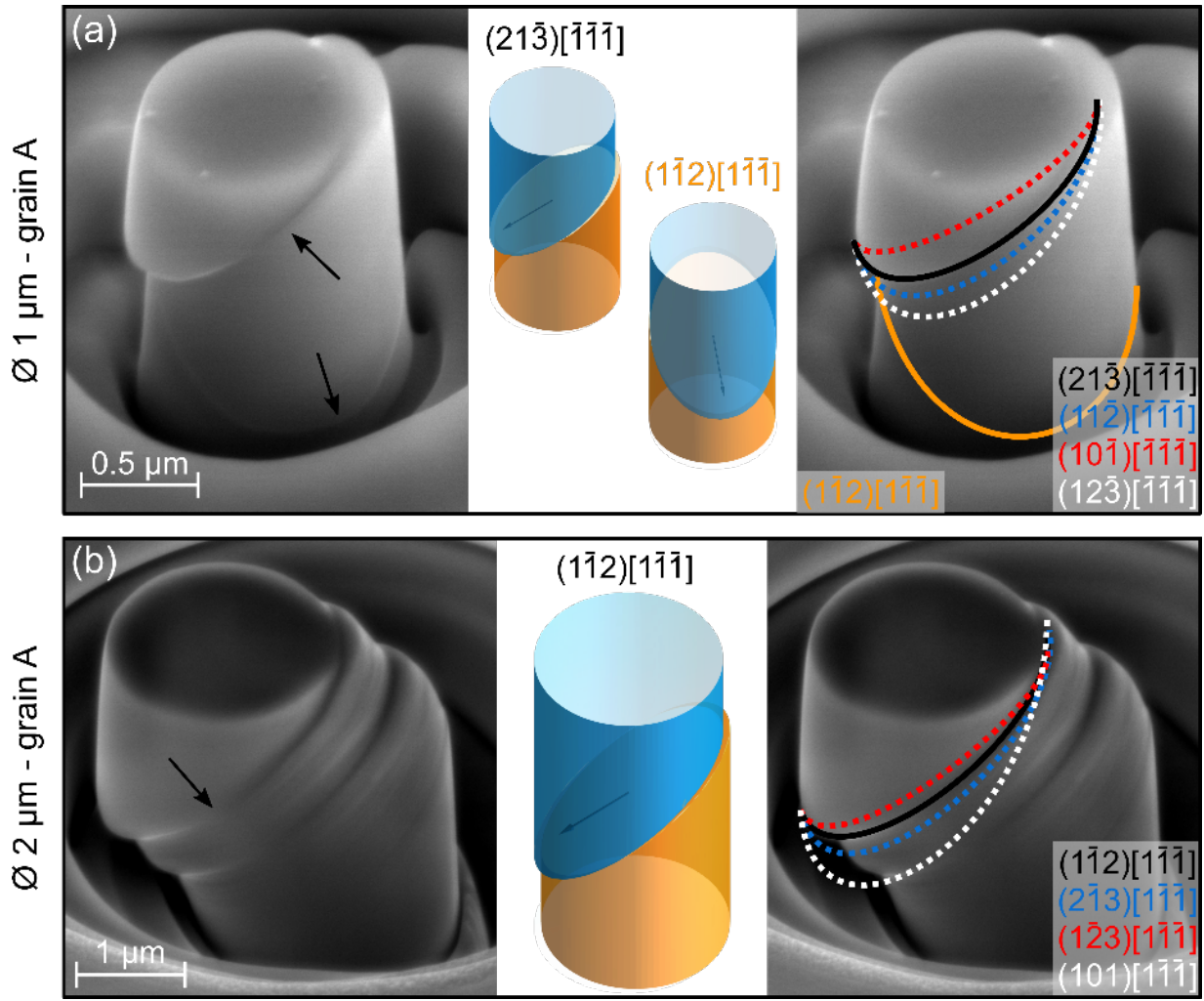

*Figure 3: Slip system analysis of two exemplary A micropillars after relative and side view rotation, (a) showing a 1 µm micropillar and (b) a 2 µm micropillar. General structure of the image: (left) SE-images with indicated slip traces (arrows), (mid) cylindrical micropillar models with identified activated slip systems, (right) SE-images with overlaid slip traces. From black to white the Schmid-factor decreases and the solid line indicates the matching slip trace.*

Figure 4 shows the results for the slip system analysis of orientation B, oriented for single slip. Figure 4 (a) shows the analysis of a 1 µm micropillar and Figure 4 (b) of a 2 µm micropillar. In both micropillars, $(3\bar{1}2)[11\bar{1}]$, with the highest Schmid factor of nearly 0.5 (supplementary material S3), is the activated slip system which is true also for all other B micropillars. Again, the micropillars exhibit fewer but clearer slip traces in smaller micropillars. More than one active slip system is not found within the same B micropillar.

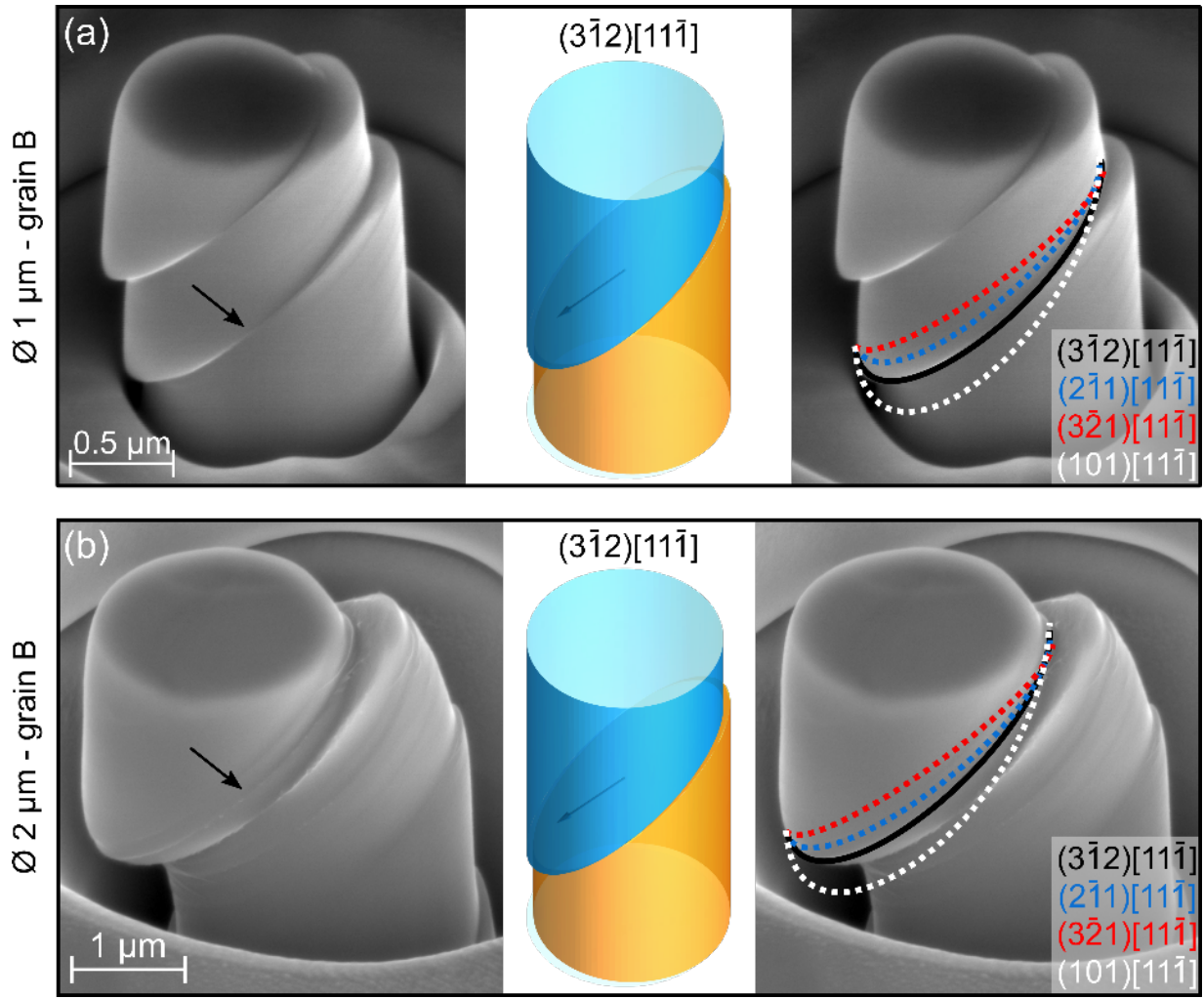

*Figure 4: Slip system analysis of two exemplary B micropillars after relative and side view rotation, (a) showing a 1 µm micropillar and (b) a 2 µm micropillar. General structure of the image: (left) SE-images with indicated slip traces (arrows), (mid) cylindrical micropillar models with identified activated slip systems, (right) SE-images with overlaid slip traces. From black to white the Schmid-factor decreases and the solid line indicates the matching slip trace.*

### 3.1.2. Bi-crystal

The slip system analysis of four bi-crystalline AB micropillars is presented in Figure 5. Some of the 2 µm micropillars with aspect ratios above 2.25 showed a bent, "S"-like behaviour (Figure 5 (c)) with many visible slip traces instead of the typical sharp slip, described as buckling by Kirchlechner et al. [16]. The resulting lateral forces are associated with difficulties in interpreting the active slip systems and the associated stresses from the engineering stress-strain curves [43] hence the need for additional verification of the slip systems via the lift-out EBSD and tilt analysis (section 4.1).

In contrast to the single-crystalline counterparts, different slip systems are activated in the bi-crystal pillars. Grain A now shows $(2\bar{1}3)[1\bar{1}\bar{1}]$ as the dominant slip system with the second highest Schmid factor (supplementary material S3). Grain B shows several active slip systems with $(3\bar{2}1)[11\bar{1}]$ (Figure 5 (a) & (c)) being the most probable, by a factor of 4, followed by $(2\bar{1}1)[11\bar{1}]$ (Figure 5 (c)) and $(\bar{1}10)[11\bar{1}]$ (Figure 5 (b)). $(3\bar{2}1)[11\bar{1}]$ has the third highest, $(2\bar{1}1)[11\bar{1}]$ the second highest and $(\bar{1}10)[11\bar{1}]$ the fifth highest Schmid factor (supplementary material S3). Some B grains show more than one visibly activated slip system. In Figure 5 (a), $(3\bar{2}1)[11\bar{1}]$ slip can be seen, indicated by the red line in the left-hand image. Apart from this (a) also shows a slip trace belonging to $(3\bar{1}2)[11\bar{1}]$ which was otherwise only found in the single-crystalline counterpart. For clarity $(3\bar{1}2)[11\bar{1}]$ is highlighted in the right-hand image, using the solid black line. Here it is worth mentioning that orientation A has between 11 and 13 slip systems with a high Schmid factor of above 0.43 and orientation B just 4-6. The fluctuation in the number of slip systems results from orientation deviations along the sample (Figure 1, supplementary material S3).

Before deformation, most grain boundaries were nearly perfectly vertical. This changed after deformation. In micropillars without pronounced buckling, the grain boundaries remained vertical until reaching an intersection with localized slip traces, from where the grain boundary was deflected at a specific angle before becoming vertical again after the slipped section (Figure 5 (a), (b)). In the buckled micropillar (Figure 5 (c)), in grain A slip traces can be found over the whole height of the pillar while in grain B slip traces are just visible in the bottom half. However, the grain boundary remains vertical until it reaches the area where slip traces are also visible in grain B, implying slip transmission is necessary for this deflection.

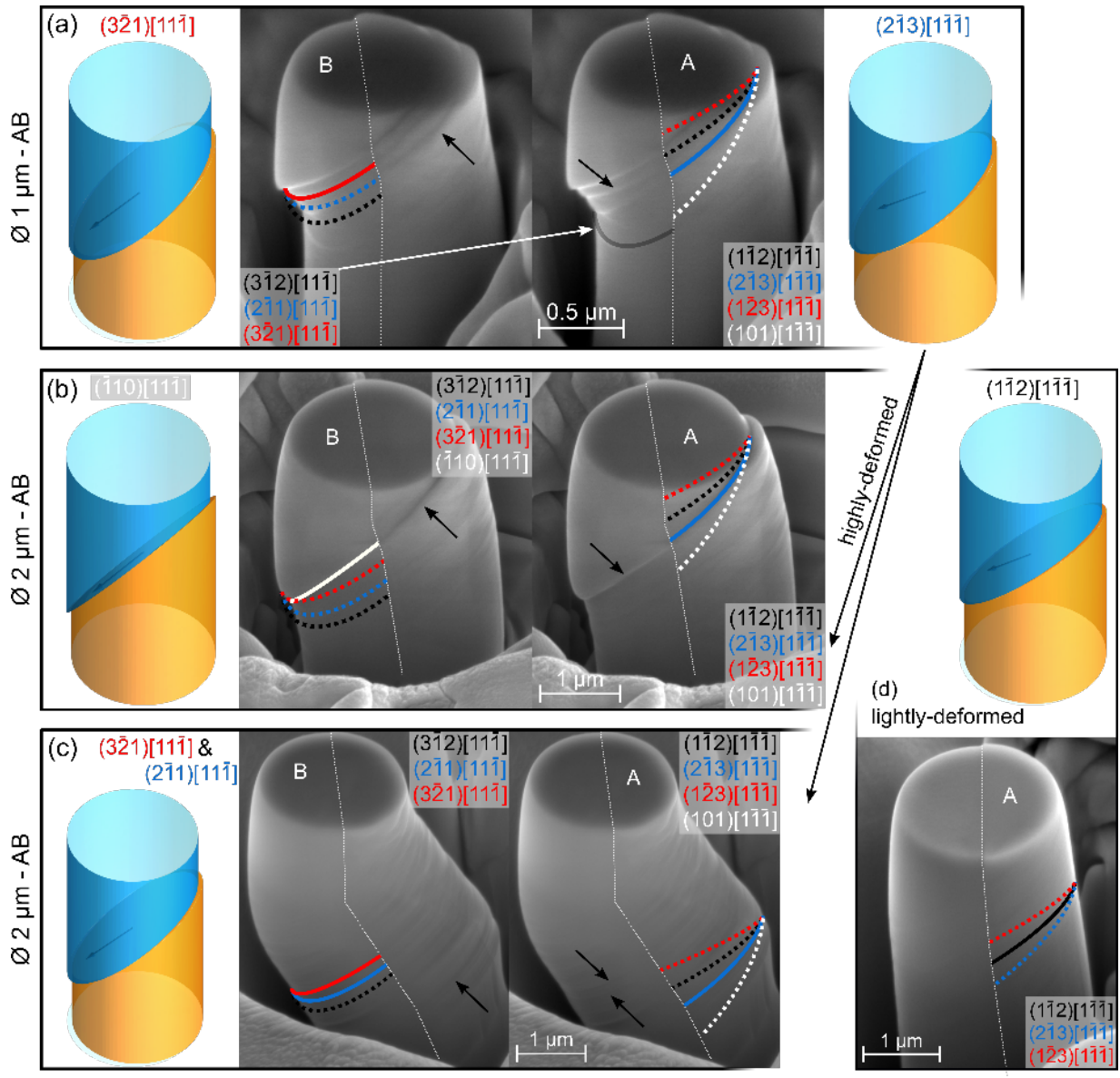

*Figure 5: Slip system analysis of four exemplary AB micropillars after relative and side view rotation, (a)-(c) showing highly-deformed (~30% total strain) 1 and 2 µm micropillars and (d) a lightly-deformed (< 9% total strain) 2 µm micropillar. General structure of the figure: (left) cylindrical micropillar models of the determined slip systems for grain B, (top-right) cylindrical micropillar model of the determined slip system for grain A, (middle) two SE-images both showing the same micropillar, one indicating the analysed slip traces for one side of the grain boundary (arrows) which are then analysed in the other picture and vice versa, from black to white/red the Schmid factor decreases*

Figure 5 (d) shows an AB micropillar, which was only deformed to a low strain. In the single-crystalline micropillars the maximum strain had no influence on the activated slip system. In contrast, in the case of the micropillar in (d) and other lightly-deformed bi-crystalline micropillars the visibly active slip system changed. The 45° tilted SE-image shows weak $(1\bar{1}2)[1\bar{1}\bar{1}]$ slip traces in grain A. Grain B has no visible slip traces.

$(1\bar{1}2)[1\bar{1}\bar{1}]$ was also the most frequently activated slip system in the single-crystalline A micropillars. Visible slip confined to grain A of the AB micropillar in addition to the slip direction from right to left in Figure 5 indicates that slip started in grain A for all AB micropillars.

Although slip systems were determined, it is important to note that the visual slip trace analysis is not straightforward and unambiguous. Uncertainties because of the subjective selection of closely related slip systems and possible misalignments between EBSD and SE-images remain. Thus, the slip system analysis will be further discussed in section 4.1.

Table 1 shows the geometrical transmission factor *m'* (see Equation 2) for different slip system combinations. A value close to 1 is equivalent to low angular differences between the slip systems and possibly easy transmission of dislocations across the grain boundary. "A & B" in Table 1 shows the theoretical *m'* value if the slip systems of the single-crystalline micropillars were also active in the bi-crystalline ones. At 0.81 this value is already quite high. Nevertheless, the values of the actually present AB slip system combinations are even higher. The most prominent combination $(2\bar{1}3)[1\bar{1}\bar{1}]$ || $(3\bar{2}1)[11\bar{1}]$ has the highest *m'* value (0.91) followed by $(2\bar{1}3)[11\bar{1}]$ || $(2\bar{1}1)[11\bar{1}]$ with 0.9, $(2\bar{1}3)[1\bar{1}\bar{1}]$ || $(3\bar{1}2)[11\bar{1}]$ with 0.88 and $(2\bar{1}3)[1\bar{1}\bar{1}]$ || $(\bar{1}\bar{1}0)[11\bar{1}]$ with 0.84. Also recorded in the last three rows are possible, but not present, combinations due to their high *m'* value of 0.87 - 0.9.

*Table 1: Observed and theoretical slip system combinations and their resulting geometrical transmission factor m'. The transmission factor for "A & B" corresponds to the theoretically expected m' for the primary slip systems found in the single-crystalline pillars.*

| reference | glide System A | glide system B | transmission factor *m'* |
|---|---|---|---|
| A & B | $(1\bar{1}2)[1\bar{1}\bar{1}]$ | $(3\bar{1}2)[11\bar{1}]$ | 0.81 |
| 1 μm - AB | $(2\bar{1}3)[1\bar{1}\bar{1}]$ | $(3\bar{2}1)[11\bar{1}]$ | 0.91 |
| 1 μm - AB | $(2\bar{1}3)[1\bar{1}\bar{1}]$ | $(3\bar{1}2)[11\bar{1}]$ | 0.88 |
| 2 μm - AB | $(2\bar{1}3)[1\bar{1}\bar{1}]$ | $(\bar{1}\bar{1}0)[11\bar{1}]$ | 0.84 |
| 2 μm - AB | $(2\bar{1}3)[11\bar{1}]$ | $(2\bar{1}1)[11\bar{1}]$ | 0.9 |
| theoretically | $(1\bar{1}2)[1\bar{1}\bar{1}]$ | $(3\bar{2}1)[11\bar{1}]$ | 0.9 |
| theoretically | $(1\bar{1}2)[1\bar{1}\bar{1}]$ | $(\bar{1}\bar{1}0)[11\bar{1}]$ | 0.87 |
| theoretically | $(1\bar{1}2)[1\bar{1}\bar{1}]$ | $(2\bar{1}1)[11\bar{1}]$ | 0.87 |

## 3.2. Mechanical Properties

Engineering stress-strain-diagrams for the single- and bi-crystalline pillars (1 to 2500 µm in diameter) are shown in Figure 6. As is common at small scales, the 1 µm micropillars in (a) show significant scatter, but trends can nevertheless be identified. AB has by far the highest yield strength (963 ± 108 MPa) followed by A (754 ± 121 MPa) and B (595 ± 42 MPa). This same trend is also seen for the 2 µm micropillars (b) but to a lesser degree. A drastic reduction in the scatter is also seen. For the 4 µm micropillars in (c), the scatter between the curves is again reduced, as is the yield strength differences between the orientations. Here the average strengths now proceed A > AB > B, although they all lie within a similar range considering the scatter in the data. In comparison, the 2500 µm pillars showed nearly no scatter or yield strength differences in pre-tests, therefore just one pillar per group was compressed with the desired strain rate. Here, orientation B also has a slightly lower yield strength and work hardening rate than A and AB.

Analysing the onset of plasticity from (d) to (a), the yield strength increases with decreasing diameter within the orientation groups and flow becomes more intermittent, which is known as the size effect [44]. It can be seen that the size effect is still present at 4 µm since the yield strength is still more than 100 MPa higher than that of the macroscopic pillars. In addition, the smaller bi-crystalline pillars generally show an increased work hardening rate compared to their single-crystalline counterparts. As with the relative strengths, this effect is negligible at 4 µm diameter, and in the macropillars, grain A is comparable to the bi-crystal with some degree of work hardening. In the supplementary material (S2) geometrical, mechanical and experimental information are summed up regarding every pillar group.

For every micropillar group, the yield stress – as determined via a 0.5% offset strain criterion – has been used to calculate the critical resolved shear stress for the primarily-observed slip system (Table 2). Taking the size effect into account these values are fairly similar regardless of the operating slip system, as would be expected for iron around room temperature, making the Schmid factor the dominating influence on the activated slip system.

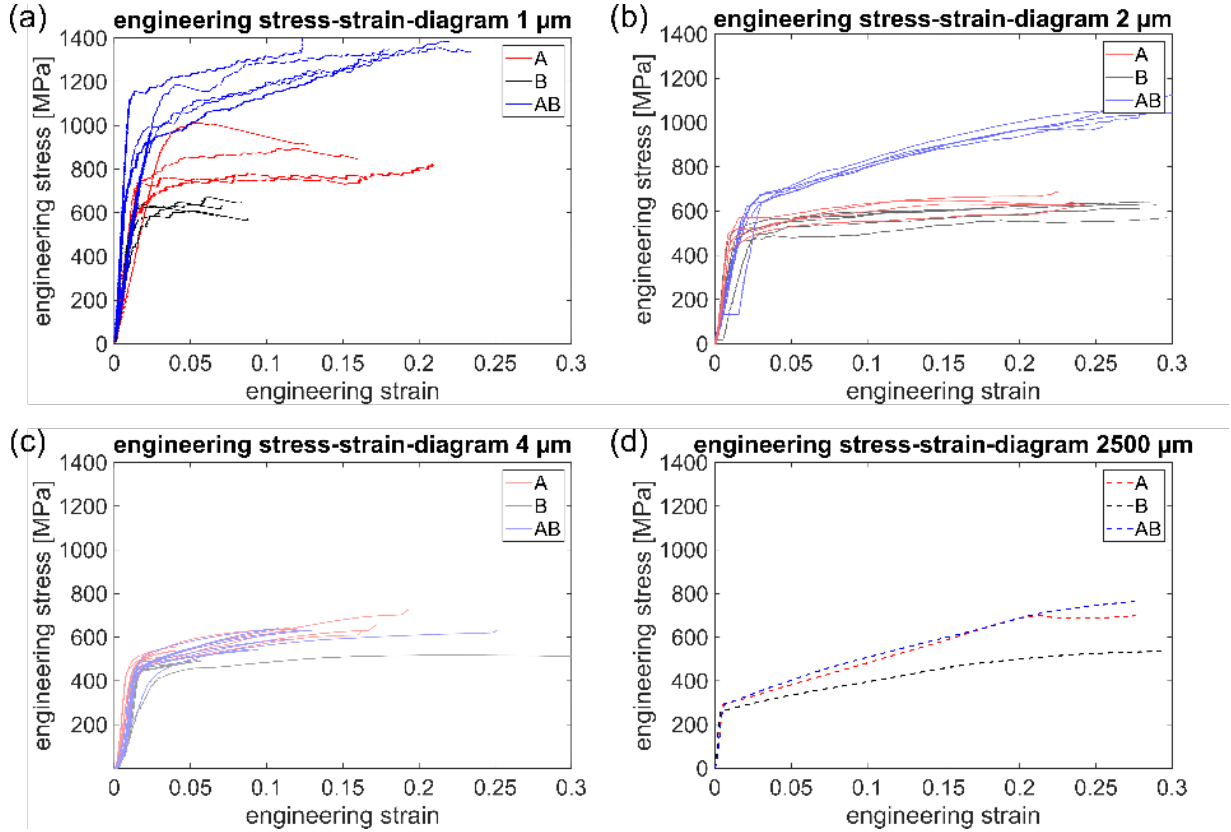

Figure 6: Engineering stress-strain-diagrams of single- (A, B) and bi-crystalline (AB) pillars with (a) 1 µm, (b) 2 µm, (c) 4 µm and (d) 2500 µm diameter.

Table 2: Sample groups, their most common slip systems with Schmid factor as well as the measured yield strength (as determined via a 0.5% offset strain criterion) and the resulting critical resolved shear stress (CRSS).

| sample | slip system | Schmid-factor | yield strength [MPa] | CRSS [MPa] |
|---|---|---|---|---|
| 1 µm - A | $(21\bar{3})[111]$ | 0.476 | 754 ± 121 | 359 ± 58 |
| 2 µm - A | $(1\bar{1}2)[1\bar{1}\bar{1}]$ | 0.498 | 504 ± 45 | 251 ± 23 |
| 4 µm - A | $(1\bar{1}2)[1\bar{1}\bar{1}]$ | 0.498 | 485 ± 7 | 242 ± 4 |
| 1 µm - B | $(3\bar{1}2)[11\bar{1}]$ | 0.5 | 595 ± 42 | 297 ± 21 |
| 2 µm - B | $(3\bar{1}2)[11\bar{1}]$ | 0.5 | 487 ± 17 | 243 ± 9 |
| 4 µm - B | $(3\bar{1}2)[11\bar{1}]$ | 0.493 | 443 ± 16 | 218 ± 8 |

## 4. Discussion

The extensive testing scheme of the presented study is important to obtain reliable results since micropillar compression is prone to statistical deviations because of the intrinsic sensitivity of small scale testing to internal and external factors [16, 43, 44]. Some intrinsic factors are the number and size statistic of dislocation sources [45-48], mechanical asymmetries during loading of bi-crystals because of the crystal structure [16, 43] or impurities. Extrinsic factors on the other hand could be a flat punch/micropillar misalignment, friction, an imperfect micropillar shape or ion beam damage.

### 4.1. Validation of slip system analysis

Although the correlation of slip lines with the crystal structure to analyse the occurring slip systems is widely used in the literature [15, 49, 50], it is still prone to errors from small misalignments between the EBSD measurements and the SE-images after deformation or because of the subjective judgement regarding the slip trace fit. The latter is especially true since the differences in terms of the angle of the intersection with the pillar surface between the slip systems are not very pronounced in BCC crystals.

Therefore, to support the results of Figure 3 to Figure 5, a second approach regarding the slip system analysis is shown in Figure 7. With the slip direction set to be horizontal, the vertical distance between the lower apex of the top face and the first slip trace was measured from opposite sides. These values, combined with the micropillar diameter, provide information about the glide plane tilt. In Figure 7 (a) and (b) the vertical distance measured on opposite sides is shown. The calculated plane tilt can then be compared with the theoretical slip plane tilt visualized in an edge-on view (90°) of a unit cell (c) or of the cylindrical micropillar model (d). Measurements on the single-crystalline A and B micropillars indicated a plane tilt of the activated slip plane of 0°. This is in good agreement with the model edge-on views (90°) of the theoretically activated slip systems determined in Figure 3 and Figure 4. The AB micropillars on the other hand show a very pronounced plane tilt in their SE-images of up to 12°. Generally, all the potential slip systems in Figure 5 show a plane tilt in the same direction within the two models, however $(1\bar{1}0)[11\bar{1}]$ has the strongest tilt, which does not match well with the calculated tilt angle. Conversely, the tilt angle of the other two found theoretical slip planes fit well. Thus most found theoretical slip systems in Figure 3 to Figure 5 are validated by this technique although $(1\bar{1}0)[11\bar{1}]$ remains a one-time exception.

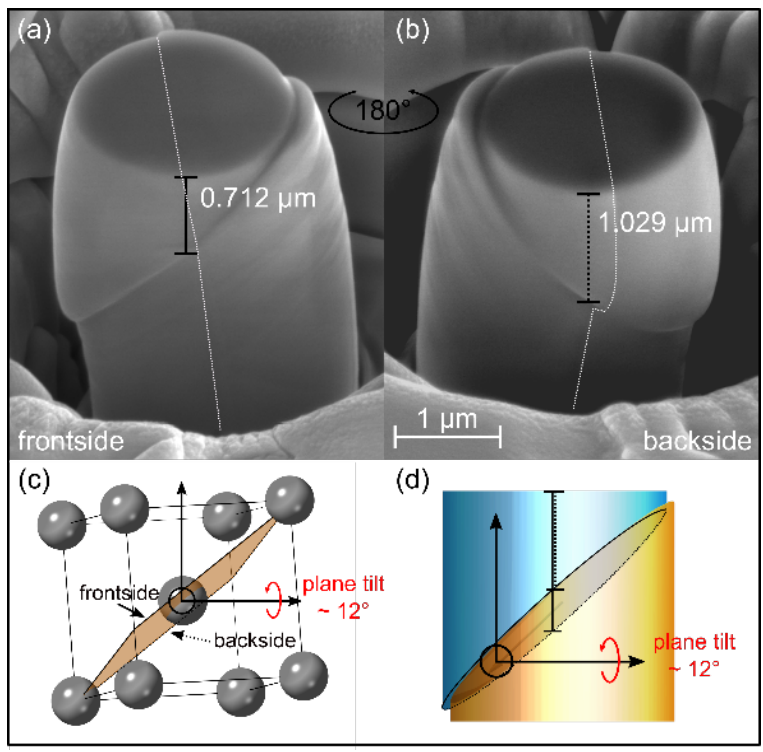

*Figure 7: (a) Exemplary SE-micropillar-image after relative and side view rotation with measured slip trace distance from the top face, (b) similar to (a) but 180° rotated around the top face normal, (c,d) visualization of AB glide plane tilt via (c) crystal maker and (d) cylindrical MATLAB model, in an edge-on view.*

In a third approach, selected micropillars were further analysed via cross section EBSD, shown in Figure 8 and Figure 10 (a)-(d). The first row (a)-(c) of Figure 8 shows the cross section EBSD results of a 2 μm – grain A micropillar after deformation with a pronounced crystal rotation of up to 11° (Figure 8 (c)). Within the crystal misorientation deviation map (Figure 8 (b)), it is possible to measure the slip plane angle indirectly in an edge-on view as dislocation storage and therefore local misorientation varies from the bulk of the pillar, and indeed a good fit with the expected slip system is seen. In addition, the slip plane normal should rotate towards the compression axis in a laterally constrained uniaxial compression test, due to rotation of the crystal. Or, in an illustration using an inverse pole figure, the crystal direction corresponding to the compression axis should move towards the pole of the expected slip plane. We use this visualisation here in Figure 8 and Figure 10, rather than a pole figure, to be able to show the different crystal orientations of the bi-crystals in one figure. The expected relative rotation is exactly seen here, as shown in Figure 8 (c) and is therefore another indication for the determined slip system $(1\bar{1}2)[1\bar{1}\bar{1}]$ in orientation A.

In Figure 8 (e), the orientation deviation from the bulk is much less pronounced for this 2 μm – grain B micropillar, due to the lower total strain of 12 % compared to 29 % for the micropillar in ((a)-(c)). Although the slip traces are quite visible, the rotation of the slip plane normal (f) and the slip plane angle (e) are less obvious. The expected slip plane angle of 48° can be seen to correlate with features of the misorientation deviation map in (e), but with large uncertainty, precluding a direct confirmation of the active slip system. To a large extent the overall rotation of the slip plane normal is still towards the compression axis, as expected. It is possible that at the beginning of deformation there was minor misalignment in the system, which was compensated by the activation of a deviant slip system, explaining the rotation axis scatter. This is supported by the SE-image (d) where undefined substructures are visible near, and on, the top face, which could be deviant slip traces. Thermal or mechanical noise combined with the high friction of the flat punch could also lead to this kind of rotation.

Regarding the bi-crystalline micropillars, Figure 10 (a)-(d) plots data from a heavily deformed micropillar. The rotated area in grain B is much larger than that in grain A. In other words, a greater number of immobile dislocations are leading to a greater rotation of the slip plane normal in grain B at the late stages of deformation. Generally, the results are not straightforward to interpret in this case. For grain A, the rotation of the slip plane normal causes the compression axis to rotate towards $(1\bar{1}2)$, like in its single-crystalline counterpart, although $(2\bar{1}3)$ is the previously determined slip plane. However, in Figure 8 (i) the spread of the rotation going towards $(2\bar{1}3)$ is higher than in Figure 8 (c). For grain B only a combination of all found slip systems would account for the measured rotation. These uncertain results could emerge from multiple active slip systems and complex interactions at the grain boundary. As expected, the analysis of the rotation of the slip plane normal is only unambiguous for single-crystals because of the much more complex stress tensor within the bi-crystal, which will be discussed further later (section 4.2).

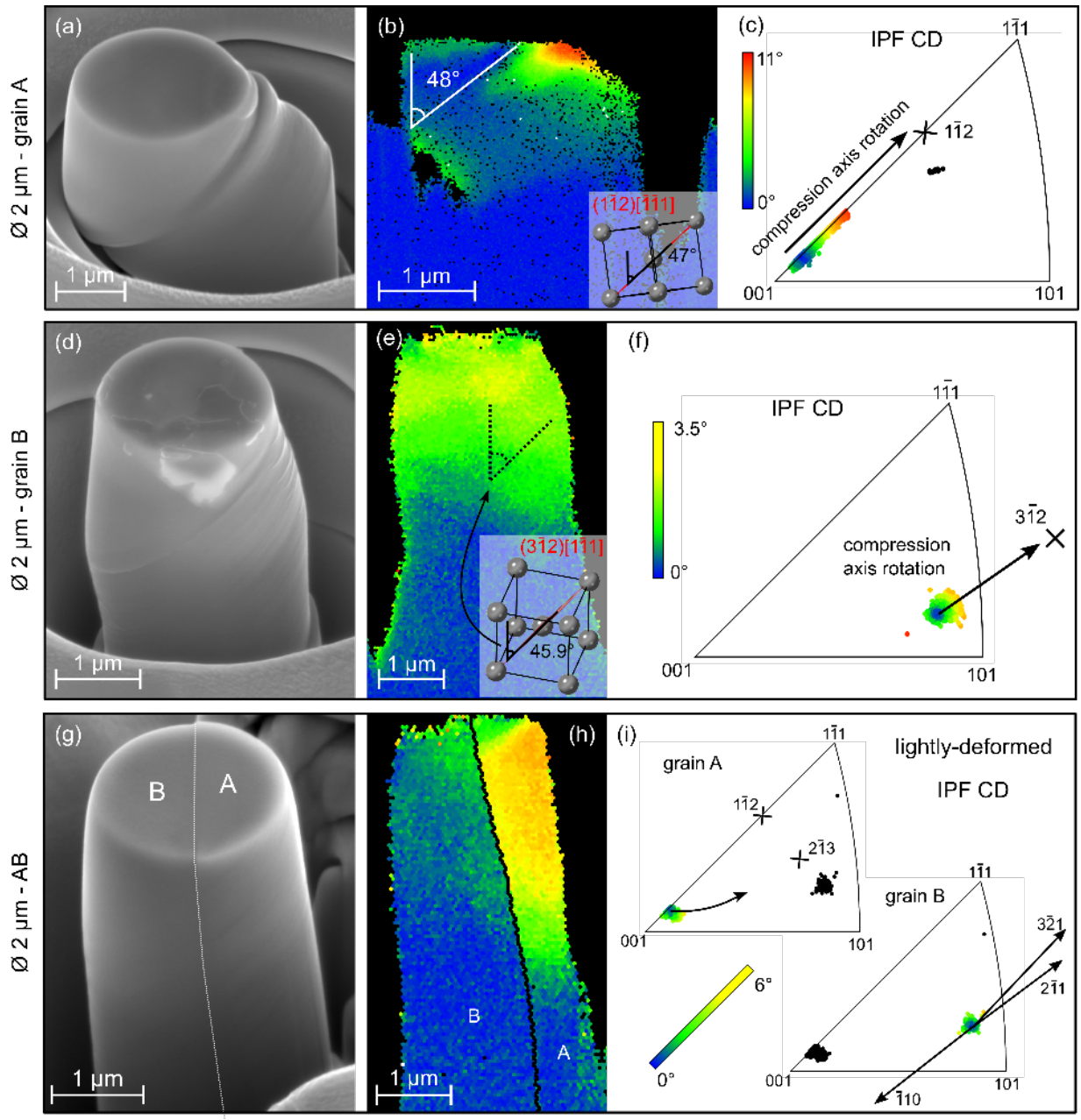

*Figure 8: Cross section EBSD results for a 2 µm – grain A, grain B and AB micropillar (lightly-deformed), (a), (d) & (g) SE-images at a 45° tilt angle, (b), (e) & (h) misorientation deviation maps with indicated slip plane angles, (c), (f) & (i) IPF triangles relative to the compression direction (CD) with indicated relative rotation of the compression axis towards the (hkl) pole of the identified slip plane, the slip plane positions and the corresponding colour codes including scales for the orientation deviation/axis rotation of (b), (e) & (h)*

In essence, all three methods come to the same conclusions with regard to the slip system analysis. Individually, the methods are susceptible to misinterpretation, but all showing similar results after individual execution strongly indicates that the determined slip systems are correct.

## 4.2. Plasticity in single- and bi-crystalline BCC micropillars

From theory {110}, {112} and {123} planes, all sharing the densest packed <111> slip direction, should be most common because of their large interplanar spacing. Above $T_c$ all these planes have been observed in BCC metals: by Tian et. al [49] in single-crystalline ferrite micropillars, as well as by Veselý [51] in macroscopic molybdenum single-crystals with {110} being the most probable. Moreover, Schneider et al. [52] assumed {112}⟨111⟩ slip for Mo micropillars at elevated temperatures. Below $T_c$ mainly {110} planes are observed, with {112} planes rarely observed and {123} planes not found [25, 53-55]. The critical temperature for iron is around room temperature [31], and in our experiments {112} and {123} planes are most frequently observed. Theoretically {123} has the highest Peierls barrier at 0 K [56], however from our results we conclude that thermal activation minimised the differences in activation energy between the mentioned slip systems (based on {110}, {112} and {123}) and other factors than the lattice resistance dominate. The observed $(1\bar{1}2)[1\bar{1}\bar{1}]$ (dominant in grain A) and $(21\bar{3})[\bar{1}\bar{1}\bar{1}]$ (dominant in grain B) slip systems have a higher Schmid factor than any system of the {110}<111> family supporting the activation of the former two, given that the values of CRSS for the slip systems are fairly comparable for micropillars of the same diameter (Table 2).

Wavy slip indicating cross slip of screw dislocation was not observed, neither in the single-crystals nor in the bi-crystals. Cross slip is otherwise regularly observed in BCC metals within the thermal regime [25]. Marichal et. al [55] concluded that wavy {112} slip traces are a visual result of the alternating activation of two {110} systems. Its absence in micropillars was explained by Schneider et al. [57] as the controlling mechanism in small scale testing being dislocation nucleation rather than propagation.

At first glance, the results of the bi-crystalline pillars are counter-intuitive. Although the considered grain boundary has a very high geometrical transmission factor for the combination of the above determined single-crystal slip systems ($m'$=0.81), high work-hardening rates (Figure 6) and different slip systems are observed. Additionally, as it can be seen from the lightly-deformed pillar (Figure 5 (d)), slip starts in grain A with the typical A single-crystalline slip system, although the compression data (Figure 6) indicates that grain A has the higher flow stress compared to grain B. This fact is additionally supported through the EBSD cross section results of a weakly deformed micropillar (Figure 8 (g)-(i)). Here most of the orientation deviation occurs in grain A and the rotation of the slip plane normal is more obvious in this grain, clearly indicating that dislocations pile up at the grain boundary and cannot overcome its structure, despite the high $m'$ value. Note that this mechanism – i.e. slip initiating in grain A followed by dislocation pile up – is seen in all pillars regardless of size.

Subsequent slip traces align at the grain boundary and continue with a slight angle deviation in the next grain (Figure 5 (a)-(c)), indicating relatively easy direct slip transmission once slip occurs on new planes. This is supported by cross section EBSD in Figure 10 (c), where the slip systems in the misorientation deviation map also align at the grain boundary with an area of low orientation deviation above. The found slip systems aligning at the grain boundary (Figure 5 (a)-(c)) in both grains are different from those in the single-crystals (Figure 3-Figure 4). Although these slip systems still have a high Schmid factor and belong to the typical BCC families, they were never found in the single-crystalline counterparts.

Summing up, slip starts in grain A on the typical single-crystalline slip systems but the dislocations cannot overcome the grain boundary (Figure 5 (d), Figure 8 (h)). It is therefore assumed that these dislocations pile up at the grain boundary, and subsequently the evolving back stresses then activate dislocation sources on the observed bi-crystalline slip systems. These are oriented for an even higher geometrical transmission factor and dislocations can then transmit through the grain boundary (Figure 5 (a)-(c), Figure 10 (c)). Thus, even small differences in the geometrical transmission factors at a grain boundary can have a huge influence on the occurring slip systems. Similar scenarios, where the grain boundary acts as an effective obstacle for dislocation movement, have already been observed in the compression of bi-crystalline nickel [21] and tin [9] micropillars as well as in simulation [7].

There could be several reasons why deformation starts in grain A. Extrinsic flaws like misalignment, load imperfections or a non-vertical grain boundary can be mostly excluded because of the high reproducibility of the observed phenomenon. An intrinsic inhomogeneous stress state on the other hand could arise from differences in the flow stresses and hardening behaviours of each grain, or differences in the Young's moduli [11, 16, 58]. As mentioned, grain A has the higher flow stress and additionally the mechanical data (Figure 6) show that the two grains have comparable work-hardening rates. Regarding the potential differences in Young's moduli, iron is known for its high elastic anisotropy [59]. In our case the loading direction of grain A is close to the "soft" [100] crystal direction and that of B is close to the "harder" [110] direction. Thus, non-symmetric resolved stresses during loading could explain the earlier onset of plasticity in grain A compared to grain B. A second explanation includes the availability of well-aligned slip systems for easy dislocation nucleation from the top face, where the pillar taper leads to the highest stresses. In grain B, the high Schmid factor directions go from the grain boundary towards the pillar side. Thus, no slip systems with high resolved shear stresses are available which can originate from the top face corner. However, for grain A there are available slip systems (e.g. the observed $(1\bar{1}2)[1\bar{1}\bar{1}]$ system) which can nucleate from this corner.

The taper of the presented micropillars is usually around the typical value of 2° [44], but in some pillars reaches almost 4.5°. The highest stress in tapered micropillars results at the top face, hence this is the reference diameter used for the calculation of stress. It is seen that the first slip lines start here, as would be expected, but importantly the observed planes do not vary in character as deformation proceeds down the pillar, nor do they vary between pillars with high (4°) and modest (2°) tapers. This high reproducibility leads to the conclusion that taper does not have an influence on which slip systems occur.

However, the stress strain data should be viewed critically. Generally, it is agreed upon that taper has a minor influence on the calculation of strain, but has a significant influence on the stress calculation and hardening behaviour [60, 61]. For example, Dehm et. al state that a taper of 3° corresponds to a maximum error of 10% in the determined yield stress depending on where the reference diameter is taken [43]. While this limits the direct transfer of the CRSS values determined here, the micropillars within this work exhibit a similar taper, and therefore the observed influence of the grain boundary, and the observed deformation mechanisms, nevertheless remain valid.

### 4.3. Size Effect

The values of $R_{p05}$ from the mechanical data in Figure 6 are plotted in Figure 9. As discussed, these show an increase in yield strength with decreasing pillar diameter, as would be expected due to the limited size and number distribution of dislocation sources in small scale samples [46-48, 62]. It can also be seen that as the diameter decreases, not only are the bi-crystal pillars stronger, the rate at which they increase in strength is faster. This matches with other work in the literature that considers the grain size as the dominating factor for bi-crystalline micropillars instead of the diameter [7, 8]. The faster rate is simply explained from the fact that the size effect usually scales with a power law, therefore the effect becomes stronger the smaller the diameter/grain size is. Here the smallest single-crystalline diameter is 0.907 µm and the smallest grain size of AB is 0.43 µm. It can also be observed that an increase in stochastic deformation behaviour and jerky flow can be seen with decreasing size, as is typical [43, 44].

When fitting a single power law to the data in Figure 9 (shown in S4 (supplementary material)) the best fit is obtained when grains A and B scale with pillar diameter and AB with the grain size ($\frac{1}{2}$ pillar diameter [8]). This also indicates that the grain boundary is an effective obstacle for the moving dislocations. Regarding the size scaling exponent, for FCC materials this is always around -0.6 [44, 63]. For BCC materials values as low as -0.25 are reported for temperatures below $T_c$ and increase to FCC-like values (e.g. -0.76) for temperatures above $T_c$ [52, 64, 65]. Schneider et. al offered three explanations for this phenomenon taking the dissociated core structure of the rate controlling screw dislocations in BCC metals into account. Firstly, kinetic pile-ups of dislocations [66], secondly, kink nucleation at the pillar-surface [67] and thirdly, that the lattice resistance component of the flow stress does not experience a size effect as the critical length scale is only of the order of nm³ (the volume of a double kink) [40, 52]. With a value of -0.37 our scaling exponent did not yet reach the FCC value, suggesting that the Peierls barrier is still relevant for the pillar strength. Thus, $T_c$ does not seem to be a sharp transition point for all aspects influenced by the lattice resistance, since we see activation of {110}, {112} and {123} slip planes, indicative of athermal behaviour, but a size effect exponent still far from this at room temperature. Furthermore, it can be seen that the size effect is not fully diminished at 4 μm since the macroscopic yield strength is nearly 200 MPa lower. With regard to the less frequent occurrence of "double slip" in the intermediate sized samples (4 μm), no explanation could be found so far.

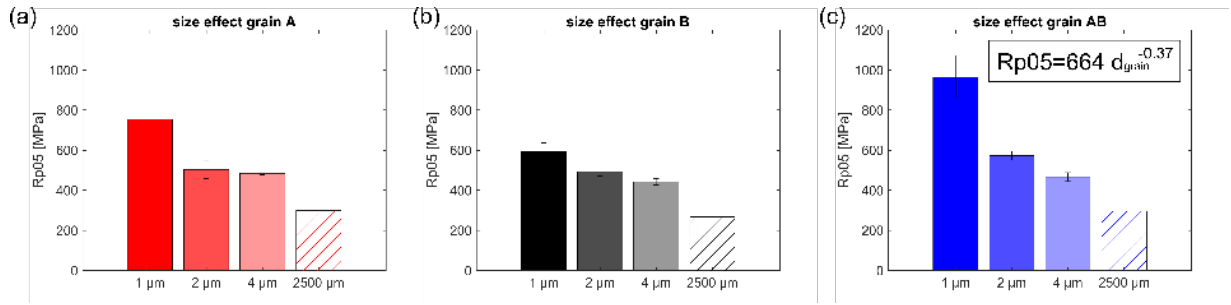

Figure 9: (a)-(c) bar charts visualizing the mean value and standard deviation of $R_{p05}$ indicating the point of onset of plastic deformation for micro- and macropillars, (a) grain A, (b) grain B, (c) AB. Additionally the size scaling law is delineated in (c) taking the respective single-crystalline data (a), (b) according to the diameter and the bi-crystalline data (c) according to the grain size ($\frac{1}{2}$ pillar diameter).

### 4.4. Scalability of the microcompression results

In order to ensure that the results are scalable from the micro- to the macroscale, not only must the size effect and differences in strain rate be taken into account, but it must also be confirmed that deformation at the boundary is not dominated by any changes in dislocation structure. In other words, the short dislocation segments in the micropillars may not be representative of those in electrical sheet with typical grain sizes of the order of at least 100 μm.

From the mechanical data alone (Figure 6), it appears that the 2500 μm pillars (especially A and AB) exhibit enhanced work hardening rates, especially when compared to the 4 μm micropillars. Under examination by optical microscopy, it is clear that dislocation activity is still confined to specific slip planes (Supplementary material, Figure S5). For grain A two slip systems are visible and for grain B just one. It is therefore hypothesised that the apparent work-hardening in grain A is due to the intersection of the two operating slip systems that are also seen in the microcompression tests. As described above, the controlling mechanism in small scale testing of single crystals is dislocation nucleation rather than propagation [57], thus explaining the absence of work-hardening in those micropillars.

To further examine deformation at the grain boundary in the macroscopic samples, additional cross section EBSD measurements were conducted on the 2500 µm - AB macropillars cut from the same grown crystal. Comparing the macroscopic IPF || CD triangles (Figure 10 (e)&(f)) to the triangles in Figure 10 (a) & (d) we note that the rotations of the slip plane normals are the same for each grain. Slip transmission cannot easily be observed in the macroscopic case (Figure 10 (g)) because of the huge quantity of slip events and the lower resolution of the EBSD map. However, the rotation of the slip plane normal is condensed around the grain boundary especially in grain B indicating that dislocations pile up. This distribution can be explained as follows: it has been shown in the micropillars that slip starts in grain A and dislocations pile up at the grain boundary leading to local stress fields. Therefore, since the grain boundary is well-aligned for slip transmission those stresses are at some point high enough for dislocations to pass the grain boundary presumably on the slip systems best-aligned for transmission, as in the micropillars. In grain B the stress levels drop with increasing distance from the grain boundary and therefore dislocation glide eventually comes to a halt partway into the grain. This subsequently leads to a high dislocation density close to the grain boundary in grain B and therefore to a high rotation of the slip plane normal. Additionally, dislocations from grain A that have been transmitted and thus are running on different slip systems in grain B can react with dislocations on the "normal" B slip planes. These reactions lead to sessile dislocations, adding up to the observed rotation of the slip plane normal.

Comparable results can be seen on the microscale. In Figure 10 (c) the rotated area in grain B is much larger than that in grain A. In other words, more immobile dislocations are leading to a greater rotation of the slip plane normal in grain B at the late stages of deformation. The similarities between the two cases therefore indicate that co-deformation of the bi-crystal does in fact occur in a consistent way at the micron and millimetre scale. This is encouraging for the use of microcompression to sample a larger number of grain boundaries within a single sheet sample compared with the need to grow a large number of macroscopic bi-crystals as utilised here to enable the bigger experimental matrix.

Finally, regarding the differences in strain rates between the microscopic and macroscopic tests (0.0001 $s^{-1}$ vs 0.01 $s^{-1}$, respectively): additional microcompression tests at comparable strain rates to the macroscopic samples (0.008 $s^{-1}$) have also been performed. The full results are not presented here as these data form part of future work in which thermal activation will be studied more comprehensively. They nevertheless did not show any differences in the deformation mechanisms in this strain rate regime.

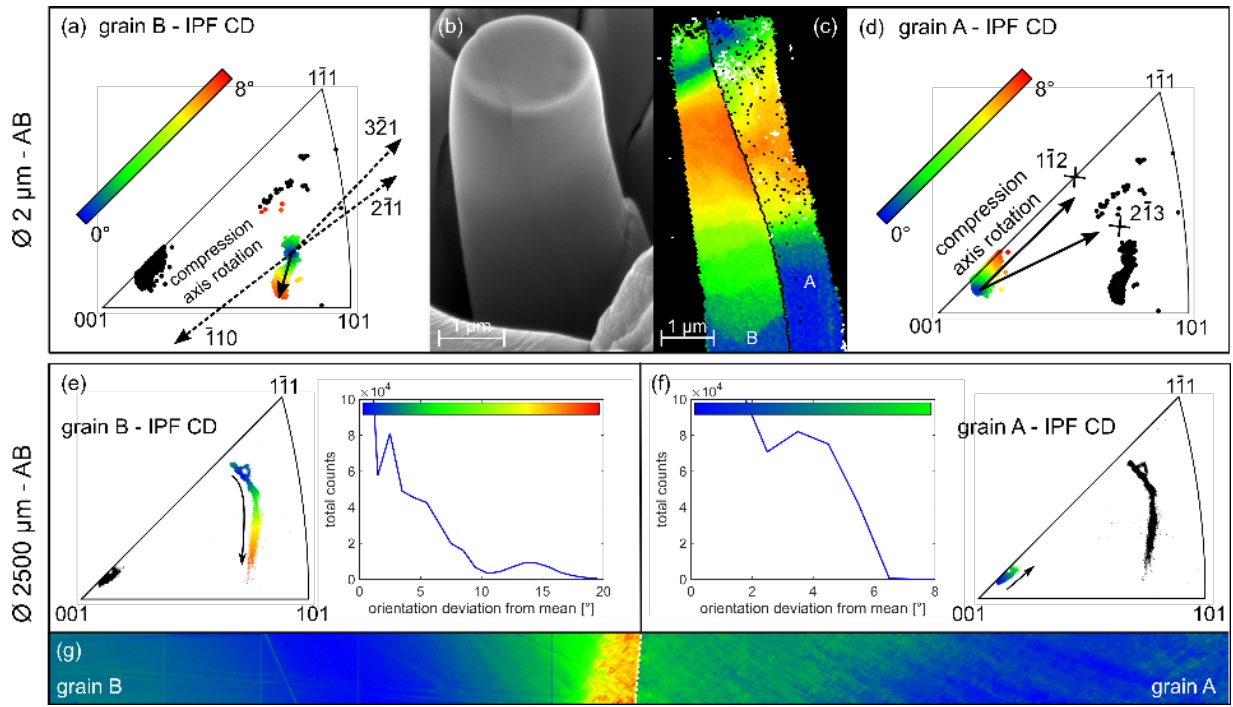

*Figure 10: Cross section EBSD results for a 2 (a)-(d) and 2500 µm – AB pillar (e)-(g), (c)&(g) coloured misorientation deviation map according to the IPF colour code of (a)|(d) and (e)|(f) both also indicating the relative rotation of the compression axis towards the (hkl) pole of the identified slip plane. Additionally (e)&(f) show graphs with the orientation deviation from the bulk for grain A and B respectively, EBSD data for the other grain is shown in black in (a),(d),(e) and (f)*

## 5. Conclusion

In the present study, we tested micropillars of different orientations and sizes as single- and bi-crystals to improve our understanding of the co-deformation at grain boundaries in Fe2.4wt.%Si, a BCC metal in which deformation at the bi- and oligo-grain level is of direct relevance to the final cutting step in electrical sheet. Due to the absence of phase transformations, this material can be grown as large single- and bi-crystals. We focused on the influence of a high angle grain boundary with a high transmission factor since few studies exist dealing with BCC bi-crystal micropillars. The following conclusions can be drawn from the obtained results:

- At room temperature $\{112\}\langle11\bar{1}\rangle$ and $\{123\}\langle11\bar{1}\rangle$ are the most activated slip systems in the two tested single-crystalline orientations indicating a similar critical resolved shear stress among the typical BCC slip system families.
- The activation of slip systems carrying the deformation and slip transmission is very sensitive to geometrical (mis)alignment of the grains at the considered grain boundary since direct slip transmission was found for slip system combinations with a transmission factor of *m'*=0.91 but not for combinations with a transmission factor of *m'*=0.81.
- Before direct slip transmission there can still be a dislocation pile up at early stages of deformation resulting in pronounced work hardening of small micropillars (1 & 2 µm)
- The size scaling exponent of -0.37 is lower than -0.6, indicating that the critical temperature is above room temperature, and that the lattice resistance therefore still contributes to the overall flow stress at this length scale (1-4 µm).
- For the considered bi-crystals the size effect scales with grain size and for single-crystals with micropillar diameter, indicating that the grain boundary is an effective obstacle for dislocation movement and source operation, as observed in FCC crystals.
- Comparable results regarding slip transmission on the micro- and macroscale prove that small scale testing can be a useful tool to learn more about the role of grain boundaries during deformation in actual applications

## Acknowledgements

This work is funded by the Deutsche Forschungsgemeinschaft (DFG, German Research Foundation) KO4603/3-2 (218259799) and carried out in the research group project "FOR 1897 Low-Loss Electrical Steel for Energy-Efficient Electrical Drives". The authors like to thank Holger Willems from the Institute of Metal Forming (RWTH Aachen University) for carrying out the macroscopic compression tests.